\documentclass[10pt,letterpaper]{article}
\usepackage[top=0.85in,left=2.75in,footskip=0.75in]{geometry}

\usepackage{changepage}

\usepackage[utf8]{inputenc}

\usepackage{textcomp,marvosym}

\usepackage{fixltx2e}

\usepackage{amsmath,amssymb}

\usepackage{cite}

\usepackage{nameref}

\usepackage{url}
\usepackage{microtype}
\DisableLigatures[f]{encoding = *, family = * }

\usepackage{rotating}

\usepackage{amsmath}
\usepackage{amssymb}
\usepackage{graphicx}
\usepackage{esint}
\usepackage{floatrow}

\usepackage{color}
\newcommand{\change}[1]{{#1}}
\newcommand{\chgtwo}[1]{{#1}}

\raggedright
\usepackage[aboveskip=1pt,labelfont=bf,labelsep=period,justification=raggedright,singlelinecheck=off]{caption}

\makeatletter
\renewcommand{\@biblabel}[1]{\quad#1.}
\makeatother

\date{}

\usepackage{lastpage,fancyhdr,graphicx}
\usepackage{epstopdf}
\fancyheadoffset[L]{2.25in}
\fancyfootoffset[L]{2.25in}
\begin{document}
\vspace*{0.35in}

\begin{flushleft}
{\Large
    \textbf\newline{Evaluation of the Intel Xeon Phi \change{7120} and NVIDIA K80 as Accelerators
for Two-Dimensional Panel Codes}
}
\newline
\\
Lukas Einkemmer\textsuperscript{1,*,\dag}
\\
\bigskip
\bf{1} Department of Mathematics, University of Innsbruck, Innsbruck, Austria
\\
\bigskip

\dag This work is  supported by the Austrian Science Fund (FWF) -- project id: P25346.

* lukas.einkemmer@uibk.ac.at 

\end{flushleft}
\section*{Abstract}
To optimize the geometry of airfoils for a specific application is an important engineering problem. In this context genetic algorithms have enjoyed some success as they are able to explore the search space without getting stuck in local optima. However, these algorithms require the computation of aerodynamic properties for a significant number of airfoil geometries. Consequently, for low-speed aerodynamics, panel methods are most often used as the inner solver.

In this paper we evaluate the performance of such an optimization algorithm on modern accelerators (more specifically, the Intel Xeon Phi 7120 and the NVIDIA K80). For that purpose, we have implemented an optimized version of the algorithm on the CPU and Xeon Phi (based on OpenMP, vectorization, and the Intel MKL library) and on the GPU (based on CUDA and the MAGMA library). We present timing results for all codes and discuss the similarities and differences between the three implementations. Overall, we observe a speedup of approximately $2.5$ for adding an Intel Xeon Phi 7120 to a dual socket workstation and a speedup between $3.4$ and $3.8$ for adding a NVIDIA K80 to a dual socket workstation.

\section{Introduction}

Numerical simulations are routinely used in applications to predict
the properties of fluid flow over a solid geometry. Such applications
range from the design and analysis of aircrafts to constructing more
efficient wind turbines. In this context, a large number of different
models and numerical methods have been developed to efficiently compute
aerodynamic quantities such as lift and drag. It is generally believed
that the compressible Navier\textendash Stokes system is able to represent
the physics that is encountered in such systems faithfully. However,
even for moderate Reynolds numbers, turbulent motion is only dissipated
at very small spatial scales. This forces an extremely fine space
discretization and renders the numerical solution of the time dependent
Navier\textendash Stokes system intractable in all but a very selective
class of applications (this approach is usually referred to as DNS
or direct numerical simulation). Consequently a hierarchy of reduced
models has been developed that are computationally more efficient.
Even though methods such as RANS (Reynolds-averaged Navier\textendash Stokes)
and LES (Large eddy simulations) are routinely employed to perform
aerodynamics simulations, these simulations can still require days
or even weeks to complete.

In this work the goal is to develop a computer program that is able
find an ideal airfoil geometry given a target function (for example,
this target could be to maximize the lift-to-drag ratio). This is
a nonlinear optimization problem as the geometry is the parameter
under consideration. In addition, the large number of maxima found
in these problems renders traditional optimization algorithms ineffective.
In recent years, genetic algorithms have enjoyed some success (see,
for example, \cite{casas2014,jones2000,gardner2003}). However, their
application yields a new computational challenge as they require the
computation of thousands or even hundred thousands of different airfoil
configuration. Consequently, even RANS or LES methods are computationally
prohibitive as the inner solver in such an optimization algorithm.

In this paper we will restrict our attention to low-speed aerodynamics.
That is, we assume that the flow under consideration is slow compared
to the speed of sound. These conditions are present in a wide range
of applications (for example, unmanned aerial vehicles and wind turbines).
Since the flow is slow compared to the speed of sound it is justified
to neglect compressible effects. In addition, we make the assumption
that the flow is irrotational. In this case the Navier\textendash Stokes
equations reduce to Laplace's equation. One should note that a direct
solution of Laplace's equation would result in a body with zero lift.
However, by imposing an additional constraint, the so-called Kutta
condition, this simple model yields very accurate results in its regime
of validity (even for lifting bodies such as airfoils, rotor blades,
or fins). In addition, many phenomenological corrections have been
developed that are able to extend the range of validity of this simplified
model considerably. 

In principle, any numerical method can be used to solve Laplace's
equation together with the Kutta condition. However, since we are
usually interested in the fluid flow outside of a solid body, so-called
panel methods (or boundary element methods) have become the standard
approach. The advantage of such a method is that only the boundary
has to be discretized. This implies that for a two-dimensional flow
only a linear system in a single dimension has to be solved (although
the corresponding matrix is no longer sparse). In addition, no error
is made by introducing an artificial boundary faraway from the dynamics
of interest. On modern computers a good implementation is able to
compute, for example, the flow over an airfoil in less than a few
tens of milliseconds (although this has not always been true in the
past). Especially in the early days of computational fluid dynamics,
performing such simulations was the only way to obtain results in
a reasonable time. As a consequence, sophisticated software packages
(such as Xfoil \cite{drela1989}) have been developed that are still
used in current aerodynamics research (see, for example, \cite{jones2000,kipouros2012,gabor2012l,fincham2015}). 

The main advantage of panel methods is that they are computationally
cheap and that fact makes them ideally suited as the inner solver
in an optimization algorithm. In addition, they are able to faithfully
reproduce the relevant aerodynamic quantities for low-speed aerodynamics
\cite{morgado2016}.

The described optimization problem lends itself well to parallelization.
As such it can potentially profit significantly from accelerators
such as graphic processing units (GPUs) or the Intel Xeon Phi. In
fact, some papers have been published that implement panel methods
on GPUs (see, for example, the work conducted in \cite{novikov2014,chabalko2014,morgenthal2014,develder2014}).
However, most of the literature focuses on the three dimensional case.
where the linear solve dominates the performance of the algorithm.
As we will see in section \ref{sec:hardware} this is not true for
the two-dimensional problem. In addition, speedups between one and
two orders of magnitude are routinely reported \cite{novikov2014,chabalko2013,chabalko2014,develder2014}.
However, since the hardware characteristics of the central processing
unit (CPU) and the graphic processing unit (GPU) do not admit such
a large difference in performance, it has to be concluded that the
performance on the GPU has been compared to a CPU implementation that
is not very well optimized. In this context it should be noted that
CPU based system now include tens of cores and thus parallelization
(and vectorization) is vital in order to obtain optimal performance
on those systems as well.

The purpose of the present work is therefore to parallelize the optimization
problem described above (of which the panel method is the computationally
most demanding part) on both traditional CPU based systems as well
as on the GPU and to compare their performance. In addition, we consider
a parallel implementation on the Intel Xeon Phi. The Xeon Phi is an
accelerator (which is added as an expansion card similar to a GPU)
based on the x86 architecture. As such this platform promises to accelerate
the computation while still enabling the use of the same development
tools (and ideally the same code) as on the CPU. For example, to parallelize
code for the Xeon Phi OpenMP is usually employed. We compare the performance
of the Xeon Phi to the implementation on the CPU and the GPU. Furthermore,
we will consider the parallelization to multiple GPUs which poses
additional challenges.

The numerical algorithm used in this paper is described in more detail
in section \ref{sec:panel}. In section \ref{sec:hardware} we then
discuss the performance characteristics of the algorithm, the hardware
used, and the general idea of the implementation. The timing results
and details of the specific implementation under consideration are
then presented in sections \ref{sec:single-gpu} (single GPU), \ref{sec:xeon-phi}
(Xeon Phi), and \ref{sec:multiple-gpu} (two GPU setup). Finally,
we conclude in section \ref{sec:conclusion}.

\section{Numerical algorithm\label{sec:panel}}

Panel methods are a type of boundary element methods. In order to
remedy the deficiency of Laplace's equation to describe the airflow
over lifting bodies, they are supplemented by the empirically derived
Kutta condition. This model, in many instances, gives a good description
of lifting flow over solid bodies for low speed aerodynamics \cite{morgado2016}.
In the following, we will limit ourselves to two-dimensional flows
over wing cross sections (so-called airfoils).

\begin{figure}[h]
\begin{centering}
\includegraphics[width=9cm]{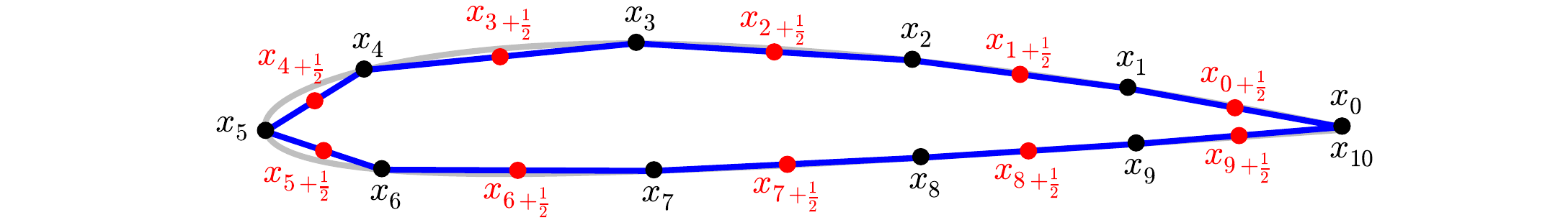}
\par\end{centering}
\caption{The discretized geometry of the NACA 2412 airfoil is shown (for the
purpose of illustration a very coarse discretization with $n=10$
is employed). The control points are shown in red and the exact geometry
is outlined in gray. \label{fig:geometry}}
\end{figure}

The geometry of the problem is given by a sequence of points $\boldsymbol{x}_{0},\boldsymbol{x}_{1},\dots,\boldsymbol{x}_{n}\in\mathbb{R}^{2}$
that represent the discretization of an airfoil $\partial\Omega$.
We assume that $\boldsymbol{x}_{0}$ is located at the trailing edge
and that $\boldsymbol{x}_{n}=\boldsymbol{x}_{0}$ holds true. This
setup is illustrated in Figure \ref{fig:geometry}.

The goal of the numerical method is to compute an approximation to
the solution of Laplace's equation in $\mathbb{R}^{2}\backslash\Omega$. \chgtwo{This solution, henceforth denoted by $\varphi$, physically represents a stream function and encodes all properties of a two-dimensional incompressible flow. For example, the velocity of the flow can be computed by $v_1 = \partial_y \varphi$ and $v_2 = -\partial_x \varphi$, where $v_1$ is the velocity in the $x$-direction and $v_2$ is the velocity in the $y$-direction. Consequently the velocity vector $\boldsymbol{v}$ is expressed as $\boldsymbol{v} = (v_1, v_2)^{\mathrm{T}}$.}

\chgtwo{Panel methods represent the solution as a superposition of translations of the  fundamental solution (which by itself is a solution of Laplace's equation everywhere except at zero)} 
\[
\phi(\boldsymbol{x})=-\frac{1}{2\pi}\log\vert\boldsymbol{x}\vert.
\]
and the global flow that is imposed far away from the airfoil.
Thus, the solution $\varphi(\boldsymbol{x})$ will be written as
\[
\varphi(\boldsymbol{x})=\int_{\partial\Omega}\gamma(\boldsymbol{s})\phi(\boldsymbol{x}-\boldsymbol{s})\,\mathrm{d}\boldsymbol{\boldsymbol{s}}+\phi_{\boldsymbol{v}}(\boldsymbol{x}),
\]
where $\gamma(\boldsymbol{s})$ is the coefficient in the superposition.
The stream function of the global flow with velocity \chgtwo{$\boldsymbol{v}=(v_{1},v_{2})^{\mathrm{T}}$}
is given by
\[
\phi_{\boldsymbol{v}}(\boldsymbol{x})=v_{1}y-v_{2}x=v_{\infty}y\cos\alpha-v_{\infty}x\sin\alpha,
\]
\chgtwo{where $v_{\infty}=\vert \boldsymbol{v} \vert$ is the speed of the global flow and} the parameter $\alpha$ is called the angle of attack \chgtwo{(note that $\boldsymbol{v} = v_{\infty}(\cos \alpha, \sin \alpha)^{\mathrm{T}}$)}. Laplace's equation
is subject to the boundary condition
\[
\varphi\vert_{\partial\Omega}=C,
\]
which enforces that no fluid can move perpendicular to the wall. \chgtwo{Note that value of $C$ will be determined as part of the numerical solution.} We
discretize this ansatz by assuming that the vortex strength $\gamma(\boldsymbol{s})$
is constant on each panel. For a panel from $\boldsymbol{x}_{i}$
to $\boldsymbol{x}_{i+1}$ \chgtwo{with vortex strength $\gamma_i$} this yields

\begin{align*}
F_{i}(\boldsymbol{x}) & =\int_{\boldsymbol{x}_{i}}^{\boldsymbol{x}_{i+1}}\gamma_{i}\phi(\boldsymbol{x}-\boldsymbol{s})\,\mathrm{d}\boldsymbol{s}\\
 & =\frac{\gamma_{i}}{2\pi}\frac{1}{\vert\boldsymbol{h}_{i}\vert}\left[\frac{1}{2}\langle\boldsymbol{x}-\boldsymbol{x}_{i},\boldsymbol{h}_{i}\rangle\log\vert\boldsymbol{x}-\boldsymbol{x}_{i}\vert^{2}\right.\\
 & \quad-\frac{1}{2}\langle\boldsymbol{x}-\boldsymbol{x}_{i+1},\boldsymbol{h}_{i}\rangle\log\vert\boldsymbol{x}-\boldsymbol{x}_{i+1}\vert^{2}\\
 & \quad-I\text{arctan2}(I,\langle\boldsymbol{x}-\boldsymbol{x}_{i},\boldsymbol{h}_{i}\rangle)\\
 & \quad\left.+I\text{arctan2}(I,\langle\boldsymbol{x}-\boldsymbol{x}_{i+1},\boldsymbol{h}_{i}\rangle)-\vert\boldsymbol{h}_{i}\vert^{2}\right]
\end{align*}
where $I=\langle\boldsymbol{h}_{i}^{\perp},\boldsymbol{x}-\boldsymbol{x}_{i}\rangle$,
$\boldsymbol{h}_{i}=\boldsymbol{x}_{i+1}-\boldsymbol{x}_{i}$, and
$\boldsymbol{h}_{i}^{\perp}$ is the outward pointing vector that
is orthogonal to $\boldsymbol{h}_{i}$ and satisfies $\vert\boldsymbol{h}_{i}^{\perp}\vert=\vert\boldsymbol{h}_{i}\vert$.
We have used $\langle\cdot,\cdot\rangle$ to denote the dot product.
The boundary condition is enforced at the control points (i.e.,~at
$\boldsymbol{x}_{i+1/2}=(\boldsymbol{x}_{i+1}+\boldsymbol{x}_{i})/2$).
This yields an underdetermined system of linear equations
\[
-\sum_{i=0}^{n-1}F_{i}(\boldsymbol{x}_{j+1/2})+C=\sum_{i=0}^{n-1}A_{ji}\gamma_{i}+C=\phi_{\boldsymbol{v}}(\boldsymbol{x}_{j+1/2})
\]
which we supplement by the Kutta condition
\[
\gamma_{0}=-\gamma_{n-1}.
\]
In stating the Kutta condition we have assumed that the variables
are ordered such that the trailing edge is located at $\boldsymbol{x}_{0}=\boldsymbol{x}_{n}$.
This, in total, gives us $n$ equations for the $n$ unknowns $\ensuremath{\gamma_{0}},\ldots,\ensuremath{\gamma_{n-2}}$
and $C$.

While the present numerical scheme yields good predictions for the
lift coefficient, it gives completely wrong results for the drag coefficient.
This is to be expected as drag is a viscous effect. However, a range
of phenomenological corrections has been developed that, for attached
flows, are able to predict the drag coefficient based on the inviscid
solution. In our code we have implemented Thwaites' method (see, for
example, \cite{tavoularis,moran1984}) in order to perform a viscosity
correction.

\begin{figure}
    \begin{minipage}[t]{.3\textwidth}
        \vspace{0pt}
        \change{
            \begin{tabular}{rrr}
                \hline 
                & \multicolumn{2}{c}{lift ($C_L$)} \tabularnewline
                \cline{2-3} 
                aoa  &  Xfoil  & Our prog. \tabularnewline
                \hline 
                0&  0.2554  &   0.2561 \\  
                1&  0.3761  &   0.3764 \\ 
                2&  0.4968  &   0.4966 \\ 
                3&  0.6173  &   0.6166 \\ 
                4&  0.7376  &   0.7364 \\ 
                5&  0.8577  &   0.8560 \\ 
                \hline 
            \end{tabular}
        }
        \end{minipage}
        \hfill
        \begin{minipage}[t]{.3\textwidth}
            \vspace{0pt}
            \includegraphics[width=4cm]{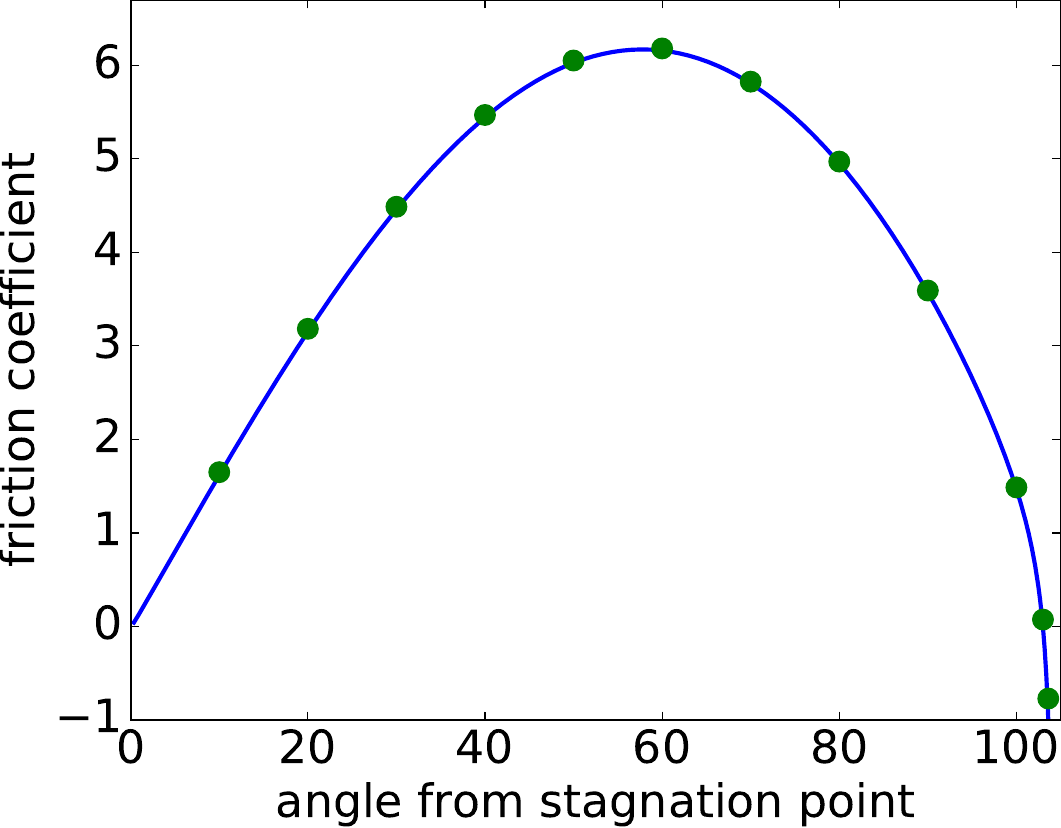}
        \end{minipage}
        \hfill
        \begin{minipage}[t]{.3\textwidth}
            \vspace{0pt}
            \change{
            \begin{tabular}{rr}
                \hline 
                \# panels & lift ($C_L$) \\
                \hline
                200  &  0.2561\\   
                250  &  0.2571\\
                300  &  0.2578\\
                500  &  0.2591\\
                1000 &  0.2600\\
                2000 &  0.2605\\
                \hline
            \end{tabular}
        }
        \end{minipage}

        \caption{\change{On the left the lift of an NACA 2412 airfoil as predicted by our program is compared to Xfoil. In the middle the drag for a circular cylinder as predicted by our program (blue line) is compared to the analytic solution of Thwaites' method (the green points correspond to some values of the analytic solution which, for example, have been tabulated in \cite{nordin1989}). On the right the dependence of the predicted lift coefficient on the number of panels used is investigated. \label{fig:validation} }}
    
\end{figure}

\change{To validate the implementation we have compared the results for the lift obtained by our program to Xfoil. As can be seen from Figure \ref{fig:validation} there is excellent agreement (the difference between the two programs is well below 1\%). Unfortunately, such a comparison is not possible for the drag as the models used for viscosity correction are different in the two programs. However, for Thwaites' method an analytic solution can be obtained for the drag over a circular cylinder. The comparison of our program with this analytic solution is shown in Figure \ref{fig:validation}. We once again observe excellent agreement. Finally, we have increased the number of panels used to discretize the airfoil. We find that it is generally sufficient to use $200$ to $300$ panels in order to obtain an error on the order of 1\%. This is certainly sufficient as neither the accuracy of the model used nor practical considerations would justify using more precision.}

\chgtwo{As has been outlined in the introduction, traditional optimization algorithms often get stuck in local minima and are thus unsuitable for the problem of interest here. Consequently the performance of a number of global search algorithms has been investigated. This includes genetic algorithms, simulated annealing, CRSA (controlled random search algorithms), etc. Among these methods genetic algorithms have been recognized as one of the best performing options (see, for example, \cite{rogalsky2000}) and have been extensively employed in a variety of applications \cite{jones2000,gardner2003,casas2014,fincham2015}.}
Therefore, we employ a genetic algorithms to perform the optimization. The first
step is to choose a parametrization of the geometry. In the language
of genetic optimization this is called the representation of the genome.
In our implementation we describe the geometry by a B-spline curve.
The location of the B-spline knot points (ordered from the trailing
edge on the upper part of the airfoil to the trailing edge on the
lower airfoil) form the representation of the genome used in the implementation.
The genetic algorithm then proceeds as follows 
\begin{enumerate}
\item Initialize a population of airfoil geometries (individuals) at random.
That is, initialize each individual by choosing the B-spline knots
at random (within reasonable bounds).
\item Evaluate the target (fitness) function for each individual using the
panel method described above.
\item Select promising individuals from the population (i.e.~individuals
with a high fitness value).
\item Combine pairs of promising individuals (parents) in order to generate
individuals for the next generation (children).
\item Perform, with a certain probability, a random mutation of a given
individual.
\item Go to 2.
\end{enumerate}
The purpose of the selection step is to favor the propagation of fitter
individuals. The rational behind this bias is that the combination
of features from two good individuals might result in an individual
with even better fitness. In our implementation we employ tournament
selection. That is, we choose $k$ individuals from the population
at random. The best individual (the individual with the highest fitness)
within that group is then selected with probability $p$. The second
best individual with probability $p(1-p)$, and so on. Two individuals,
selected in the manner described, are then combined into two children
by a crossover operation. The crossover is performed by choosing (at
random) a position in the genome (the list of B-spline coefficients)
and all coefficients prior to that point are taken from the first
parent while all coefficients starting at that point are taken from
the second parent (this is usually referred to as one-point crossover).
By reversing the order of the two parents, we obtain a second child.
This procedure is repeated until the new generation has the desired
number of individuals. The final step in the algorithm is then to
perform so-called mutations. That is, for each individual there is
a certain probability that we perturb one of its B-spline coefficients.
Mutation is crucial in order to prevent the premature convergence
of the algorithm. If the probability of mutation is too low, the algorithm
can easily get stuck in a local maximum (which we strive to avoid).
For more details on genetic algorithms we refer the reader to \cite{poli2008}.

\begin{figure*}

\begin{center}
\includegraphics[width=12cm]{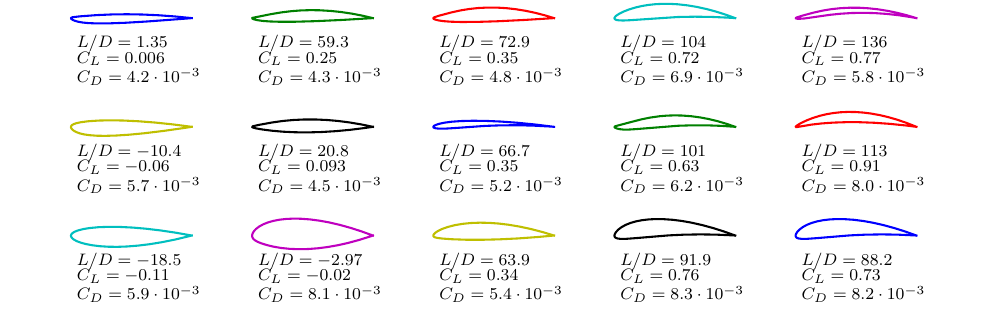}
\par\end{center}

\caption{Three airfoils for generation $1$,$2$,$3$,$6$, and $7$ of the genetic optimization algorithm are shown. The algorithm proceeds from the left to the right and each column represents a distinct generation. We show the best classes of airfoils (according to the lift-to-drag ratio) for a specific generation. The population size is equal to 1000.\label{fig:airfoils}}
\end{figure*}

\begin{figure*}

\begin{center}
\includegraphics[width=6cm]{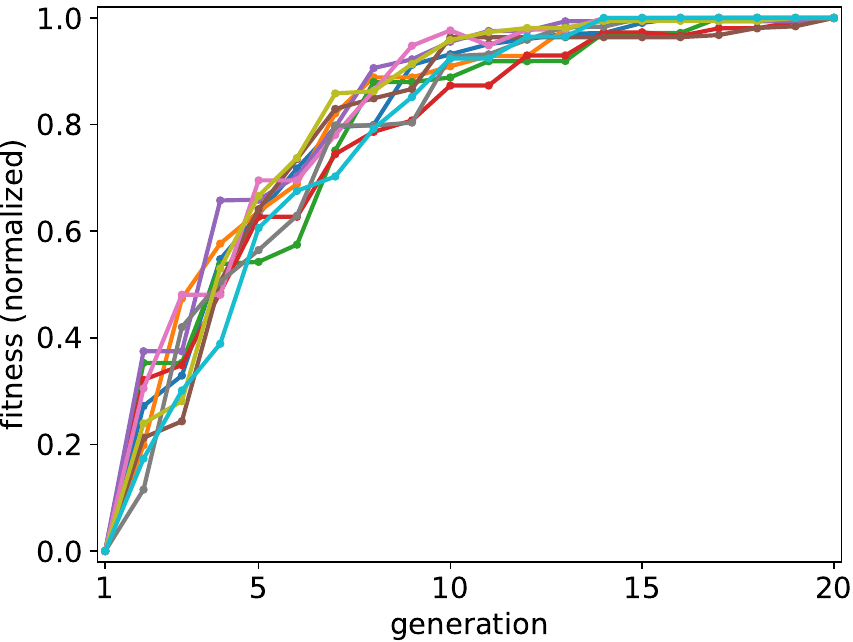}
\end{center}
\caption{\chgtwo{The fitness of the population (normalized to the overall best individual) as a function of the generation in the genetic algorithm is shown for ten different (random) initial configurations.} \label{fig:convergence}}

\end{figure*}

In Figure \ref{fig:airfoils} the evolution of the optimization algorithm
is shown. In this simulation the fitness function
is proportional to the lift-to-drag ratio at zero angle of attack.
The lift and drag coefficients stated in the Figure are computed using
Xfoil (as opposed to using the output of our simulation). This is
done in order to validate that our code performs as expected.
\chgtwo{In addition, we have investigated the convergence of the genetic
algorithm as a function of the number of generations computed. The result
is shown in Figure \ref{fig:convergence}}.

\section{Computational considerations\label{sec:hardware}}

The numerical implementation of the above algorithm requires two parts
of significant computational effort. First, the system of linear equations
has to be assembled which requires $\mathcal{O}\left(n^{2}\right)$
operations but involves the (expensive) evaluation of two logarithms
and two $\text{arctan2}$ functions for each panel. Second, the solution
of the linear system of equations is usually done by an LU decomposition
and thus involves $\frac{2}{3}n^{3}$ operations. In practice $n$
is often between $100$ and $300$. In this regime both parts of the
algorithm require substantial computational effort.

In the following we will consider the CPU, Xeon Phi, and GPU configuration
listed in Table \ref{tab:hw-config}. These will be used for all the
numerical simulations and all the performance measurements conducted
in this paper. The corresponding (peak) performance characteristics
with respect to single and double precision arithmetics and the theoretical
attainable memory bandwidth are listed in Table \ref{tab:hw-config}.
All of these components are part of a single dual socket workstation.
\begin{table}[h]
\caption{Hardware characteristics of the dual socket workstation used in the
numerical simulations. Peak arithmetic performance for single and
double precision and the theoretically attainable memory bandwidth
are listed.\label{tab:hw-config}}

\centering{}%
\begin{tabular}{rrrrrr}
\hline 
 &  & \multicolumn{2}{c}{TFlops/s} &  & \tabularnewline
\cline{3-4} 
 &  & double & single &  & GB/s\tabularnewline
\hline 
1x E5-2630 v3 &  & 0.3 & 0.6 &  & 59\tabularnewline
2x E5-2630 v3 &  & 0.6 & 1.2 &  & 59\tabularnewline
1x Xeon Phi 7120 &  & 1.2 & 2.4 &  & 352\tabularnewline
0.5x K80 &  & 1.5 & 4.4 &  & 240\tabularnewline
1x K80 &  & 2.9 & 8.7 &  & 480\tabularnewline
\hline 
\end{tabular}
\end{table}

Some fairly representative single and double precision timing results
are collected in Table \ref{tab:cpu-vs-all}. These results point
the clear picture that on the CPU assembling the matrix is between
$2.5$ and $3.5$ times more expensive compared to solving the resulting
linear systems. Thus, on the CPU the assembly actually dictates the
performance of the algorithm to a large extend. This situation is
reversed for both the Xeon Phi \change{7120} and the K80 GPU. For the Xeon Phi \change{7120} the
assembly step is by approximately a factor of two faster compared
to the two CPUs. Since assembly is an extremely compute bound problem
and giving the similarities of the two architectures, this gain is
expected based on the factor of two difference in the theoretical
arithmetic performance. On the other hand, one half\footnote{Note that the NVIDIA K80 is a single expansion card that includes
two identical GPUs each with its own separate memory. Thus, using
one-half of the K80 means that we use one of the two GPUs present
in the system.} of the K80 outperforms the same two CPUs by a factor of approximately
$5$ and the \change{Xeon Phi 7120} by approximately a factor of $3$. Note that the
GPU architecture includes a number of so-called multi-function units
(MUFU) per streaming mutiprocessor. These are used to accelerate the
computation of certain transcendental functions. Let us emphasize
that double precision support of the multi-function units is limited.
However, double precision support for the reciprocal (which is used
in the assembly code generated for both the log and the atan2 function)
is available. 

\begin{table}[h]

\caption{Time in seconds that is required to perform the assembly and linear
solver step in our panel code. In the simulation 4000 candidate solutions
(airfoil geometries) are optimized using a genetic algorithm with
10 generations. Each geometry is discretized using 200 points. For
the linear solve we use the Intel MKL 2015 library (on the CPU and
Intel Xeon Phi \change{7120}) and the MAGMA 1.7.0 linear algebra library on the
\change{NVIDIA K80}. All measured times are in units of seconds. \label{tab:cpu-vs-all}}

\begin{centering}
\textbf{single precision}

\begin{tabular}{lrrr} \hline
              &   Assembly &   Solve &   Total \\ \hline
E5-2630 v3    &       4.93 &    1.68 &    6.61 \\
2x E5-2630 v3 &       2.70 &    1.00 &    3.70 \\
Phi 7120      &       1.35 &    3.60  &    4.95 \\
0.5x K80      &       0.46 &    3.70  &    4.16 \\
\hline \end{tabular}
\par\end{centering}
\begin{centering}
\bigskip{}
\par\end{centering}
\centering{}\textbf{double precision}

\begin{tabular}{lrrr} \hline
              &   Assembly &   Solve &   Total \\ \hline
E5-2630 v3    &       9.26 &    2.80  &   12.05 \\
2x E5-2630 v3 &       5.11 &    1.91 &    7.01 \\
Phi 7120      &       2.69 &    4.72 &    7.41 \\
0.5x K80      &       0.79 &    4.42 &    5.21  \\
\hline \end{tabular}
\end{table}

The performance of the linear solver is relatively
poor on both the \change{Xeon Phi 7120} as well as on the \change{NVIDIA K80}. Note that in our
application we are not interested in solving large linear systems
(for which both of these libraries provide excellent performance)
but in solving a large number of relatively small linear systems.
\chgtwo{In this situation the linear solve is not necessarily compute bound (this is particularly true on architectures with a high flop/byte ratio). In addition, the
irregular memory access patterns encountered in this algorithm also favor systems with more elaborate caches. Let us note that it might be possible to 
improve the performance of the linear solve on the Xeon Phi.}
In fact, some research has already been conducted
in this direction (see, for example, \cite{dong2014,vladimirov2015}).
The same is presumably true for the GPU.

It thus seems that neither the CPU nor accelerators are ideally suited
for the problem under consideration. However, since the accelerators
are very efficient in the assembly step and the CPUs are very efficient
in the linear solve step, the hope is that a hybrid algorithm that
uses both platforms can succeed in obtaining a significant speedup
compared to a CPU only implementation. The difficulty in this approach
is that a large amount of data has to be transferred over the (relatively)
slow PCIe bus. In the problem under consideration this means that
all the assembled matrices have to be transferred from the accelerator
to the CPU. Clearly, if such a scheme is to be successful some strategy
has to be employed to mitigate this communication overhead. To present
an efficient implementation and the corresponding benchmark results
for both the Intel Xeon Phi \change{7120} and for the NVIDIA K80 is the purpose
of the remainder of this paper.

To conclude this section let us mention the development tools used
on the respective platforms. On the CPU and the Xeon Phi we employ
the Intel C++ compiler and, in order to perform the parallelization,
OpenMP.
To solve the linear system on the CPU the Intel MKL library
is used (which provides highly optimized LAPACK routines). For the
GPU implementation we employ the CUDA framework and, for the multiple
GPU implementation, the MAGMA linear algebra library.

\section{GPU implementation\label{sec:single-gpu}}

\begin{figure*}

\begin{center}
\includegraphics[width=10cm]{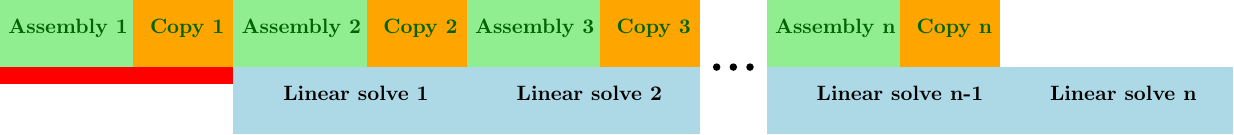}
\par\end{center}

\caption{This figure shows a communication hiding pattern that interleaves the assembly (green; on the GPU) and copy (orange; data transfer from the GPU to the CPU) with the linear solve (blue; on the CPU). The red areas constitute the remaining overhead that decreases as we divide our problem into more and more slices.\label{fig:GPU-interleave1}}
\end{figure*}

In this section we will consider an implementation where the assembly
of the matrix is conducted exclusively on the GPU and the linear solves
are performed exclusively on the CPU. This requires the transfer of
a large number of matrices in each step from the GPU to the CPU. Timing
results indicate that the run time of the assembly step (on the GPU)
together with the required transfer of data (from the CPU to the GPU)
is comparable or smaller than the time it takes to perform the linear
solve (on the CPU). Thus, to hide the communication overhead, we interleave
the assembly and transfer operation with the linear solves on the
CPU. \chgtwo{This is possible since, in principle, the assembly step can be computed independently for each individual in the population. There is, however, a computational advantage in aggregating multiple such operations together in a single slice. Therefore, we divide the population into (usually between $5$ and $20$) subpopulation. Each of these slices of the population is then assembled and send to the CPU. While the CPU is conducting the linear solve another slice is assembled on the GPU.} This approach is illustrated in Figure \ref{fig:GPU-interleave1}.
\chgtwo{In the implementation CUDA streams are used to asynchronously compute on the GPU as well as to asynchronously transfer data from the GPU to the CPU.}
It is also possible to interleave the assembly and copy operations.
However, for the GPU we found that this does not result in an increase
in performance. Thus, for the remainder of this section we will restrict
ourselves to the two-way interleave scheme illustrated in Figure \ref{fig:GPU-interleave1}.

Note that the overhead of this approach decreases as we increase the
number of slices our problem is partitioned into. However, since the
individual problems become smaller and smaller, overhead inherent
in the different parts of the algorithm becomes more pronounced. Therefore,
a compromise has to be made. In general, between $10$ and $20$ slices
seems to yield near optimal performance in most circumstances.

The timing results are given in Table \ref{tab:GPU-timing}. We observe
a speedup of $3$ (single precision) and $2.9$ (double precision)
for adding a single K80 to the dual socket workstation. Although even
a naive implementation (i.e.,~doing the assembly, the data transfer,
and the linear solve in sequential order) results in some speedup,
the communication hiding scheme employed contributes significantly
to the performance of the implementation. In the case of a single
socket workstation the observed speedup is approximately $3.6$ (single
precision) and $4.0$ (double precision).

\begin{table}
\begin{centering}
    \caption{Timing results for the hybrid algorithm (\change{one-half NVIDIA K80}+CPU) illustrated in Figure
\ref{fig:GPU-interleave1}. The wall time (W), the time required to
assemble the system (A), the time required for the linear solves (L)
    and the overhead due to offloading to the GPU (O) are shown. \change{Note that for the GPU implementation the time required by the linear solve (which is done on the CPU) always dominates the total runtime. Thus, we have $\text{W} = \text{L}+\text{O}$.}
The number of slices that yield the optimal run time are shown in bold. All measured
	times are in units of seconds. \change{In addition, the standard deviation determined from $20$ repetitions of the simulation is shown next to the wall time.} \label{tab:GPU-timing}}
\par\end{centering}
\begin{centering}
\textbf{single precision }
\par\end{centering}
\begin{centering}
\begin{tabular}{lllllll} \hline  Hardware   & slices   & W     & L    & O  & A     & speedup   \\ 
\hline
	CPU        & --       & 6.61\change{$\pm$0.01} & 1.68 & --   & 4.93 & -- \\
	GPU+CPU    & 1        & 2.68\change{$\pm$0.05} & 1.66 & 1.02 & 0.46 & 2.46      \\
						& 5        & 1.95\change{$\pm$0.02} & 1.71 & 0.24 & 0.46 & 3.39      \\
						& 10       & 1.90\change{$\pm$0.02} & 1.75 & 0.15 & 0.46 & 3.48      \\
						& 15       & 1.86\change{$\pm$0.02} & 1.75 & 0.12 & 0.47 & 3.55      \\
					 & \textbf{20}       & \textbf{1.84\change{$\pm$0.02}} & \textbf{1.73} & \textbf{0.10} & \textbf{0.47} & \textbf{3.60}      \\
             &          &      &      &      &      &         \\

	2xCPU      & --       & 3.70\change{$\pm$0.03} & 1.00 & --   & 2.70  & --        \\
	GPU+2xCPU  & 1        & 2.06\change{$\pm$0.03} & 1.03 & 1.03 & 0.46  & 1.80      \\
						& 5        & 1.32\change{$\pm$0.02} & 1.08 & 0.24 & 0.46  & 2.81      \\
						& 10       & 1.24\change{$\pm$0.03} & 1.10 & 0.15 & 0.46  & 2.97      \\
						& 15       & 1.24\change{$\pm$0.02} & 1.12 & 0.12 & 0.46  & 2.99      \\
						& \textbf{20}      & \textbf{1.22\change{$\pm$0.02}} & \textbf{1.12} & \textbf{0.10}  & \textbf{0.47} & \textbf{3.03}      \\
\hline
\end{tabular}
\par\end{centering}
\begin{centering}
\bigskip{}
\par\end{centering}
\begin{centering}
\textbf{double precision}
\par\end{centering}
\centering{}\begin{tabular}{lllllll} \hline  Hardware   & slices   & W    & L    & O    & A     & speedup   \\
\hline
	CPU        & --       & 12.05\change{$\pm$0.01} & 2.80 & --    & 9.26 & --   \\
	GPU+CPU    & 1        & 4.80\change{$\pm$0.04} & 2.90 & 1.90  & 0.77 & 2.51  \\
						& 5        & 3.39\change{$\pm$0.06} & 2.93 & 0.47  & 0.77 & 3.55  \\
						& 10       & 3.09\change{$\pm$0.07} & 2.82 & 0.27  & 0.77 & 3.89  \\
						& \textbf{15}       & \textbf{2.99\change{$\pm$0.09} } & \textbf{2.77} & \textbf{0.22} & \textbf{0.78} & \textbf{4.03}  \\
						& 20       & 3.08\change{$\pm$0.07 } & 2.88 & 0.20  & 0.78 & 3.91  \\ 
 		   &          &       &      &      &      &       \\

	2xCPU      & --       & 7.01\change{$\pm$0.01} & 1.91 & --   & 5.11 & --    \\
	GPU+2xCPU  & 1        & 3.93\change{$\pm$0.03} & 2.02 & 1.91 & 0.77 & 1.79  \\
						& 5        & 2.59\change{$\pm$0.03} & 2.12 & 0.47 & 0.77 & 2.70  \\
						& 10       & 2.45\change{$\pm$0.02} & 2.15 & 0.29 & 0.78 & 2.86 \\
						& \textbf{15}       & \textbf{2.44\change{$\pm$0.02} } & \textbf{2.20} & \textbf{0.24} & \textbf{0.78} & \textbf{2.88} \\
						& 20       & 2.44\change{$\pm$0.03 } & 2.23 & 0.21  & 0.78 & 2.87  \\ 
\hline \end{tabular}
\end{table}
The overhead in this implementation can be partitioned into two parts:
\begin{itemize}
\item As we partition our problem into more and more slices the performance
of the linear solver on the CPU decreases. This is a consequence of
the overhead required for the asynchronous data transfer to the GPU
as well as the overhead that is incurred in decreasing the batch size
for the linear solver. In the numerical simulations conducted here
this overhead is on the order of 10\%.
\item There is an inherent overhead in the interleave scheme (see the red
area in Figure \ref{fig:GPU-interleave1}). This overhead decreases
as we increase the number of slices.
\end{itemize}
Assuming instantaneous data transfer, the optimal run time of our
hybrid implementation is equal to the time for the linear solver.
Our implementation is, depending on the configuration, within $5$\%
(double precision, single socket) to $25$\% (double precision, dual
socket) of that value.

\section{Intel Xeon Phi implementation\label{sec:xeon-phi}}

In essence the implementation on the Xeon Phi is similar to the GPU
implementation. However, there are two major differences. First, due
to the 512 bit wide vector units, vectorization is extremely important
to obtain good performance on the Xeon Phi. In order to enable the
compiler to generate efficient code for the assembly step, we have
added \texttt{\_\_restrict} and \texttt{const} keywords to our \chgtwo{computational kernels.}
\chgtwo{This is rather straightforward to do as the computational kernels are implemented
using simple data structures and abstractions  are only build on top of that layer.}
We have used the vectorization report of the Intel C compiler to check
that the compiler has indeed sufficient information to vectorize the
time intensive portions of our algorithm. This has to be contrasted
with the GPU implementation of the assembly step which is relatively
straightforward (neither warp divergence nor coalesced memory access
is a major concern in this application). Note, however, that the code
for the Intel Xeon Phi is essentially identical to the optimized code
for the CPU. 

\begin{figure*}

\begin{center}
\includegraphics[width=10cm]{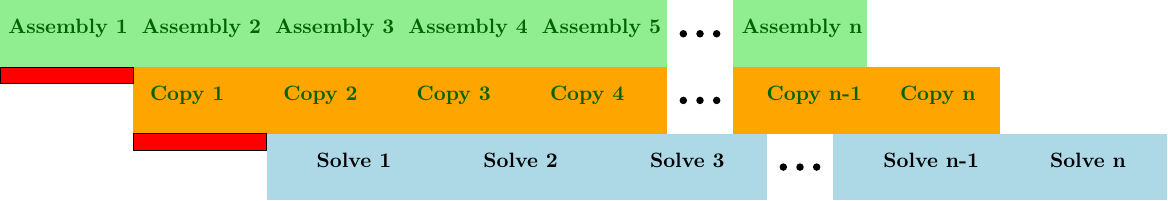}
\par\end{center}

\caption{This figure shows a communication hiding pattern that interleaves the assembly (green; on the Xeon Phi), the copy (orange; data transfer from the Xeon Phi to the CPU), and the linear solve (blue; on the CPU). The red areas constitute the remaining overhead that decreases as we divide our problem into smaller and smaller slices.\label{fig:GPU-interleaved-2}}
\end{figure*}

Second, since the assembly step takes significantly longer on the
Xeon Phi \change{7120} compared to the \change{NVIDIA K80}, it is no longer true that assembly (on
the Xeon Phi) together with data transfer (from the Xeon Phi to the
CPU) consumes less time than the linear solver (on the CPU). Thus,
in order to obtain good performance we have to interleave all three
operations as shown in Figure \ref{fig:GPU-interleaved-2}.
\chgtwo{All data transfer operations to and from the Xeon Phi are explicitly handled 
in the code. If this is not done a significant performance penalty is incurred. In order to avoid any
overhead due to the quite expensive memory allocation on the Xeon Phi, the memory
required for the computation is only allocated once (at the beginning of the simulation).}

The timing results for the Xeon Phi \change{7120} are given in Table \ref{tab:timing-xeon-phi}.
We observe a speedup of approximately $2.5$ (for both single and
double precision) for adding a single Xeon Phi 7120 to the dual socket
workstation. On the other hand, for a single socket workstation the
observed speedup is approximately $3.2$ (single precision) and $3.5$
(double precision).

\begin{table}[h]
\caption{Timing results for the hybrid algorithm (Xeon Phi \change{7120}+CPU) illustrated
in Figure \ref{fig:GPU-interleaved-2}. The wall time (W), the time
required to assemble the system (A), the time required for the linear
solves (L) and the overhead due to offloading to the Phi (O)
is shown. \change{Note that the overhead is defined such that $\text{W}=\text{L}+\text{O}$.}
The number of slices that yield the optimal run time are
highlighted in bold in the table. All measured times are in units
of seconds. \change{In addition, the standard deviation determined from $20$ repetitions of the simulation is shown next to the wall time.}\label{tab:timing-xeon-phi}}

\begin{centering}
\textbf{single precision }
\par\end{centering}
\begin{centering}
\begin{tabular}{lllllll} \hline  Hardware   & slices   & W    & L    & O   & A     & speedup   \\
\hline  
	CPU        & --       & 6.61\change{$\pm$0.01} & 1.68 & --   & 4.93 & --        \\
	Phi+CPU    & 1        & 3.67\change{$\pm$0.02} & 1.69 & 1.98 & \change{0.88} & 1.80      \\
						& 5        & 2.31\change{$\pm$0.02} & 1.71 & 0.59 & \change{0.99} & 2.87      \\
						& 10       & 2.12\change{$\pm$0.03} & 1.69 & 0.43 & \change{1.04} & 3.12      \\
						& 15       & 2.15\change{$\pm$0.03} & 1.75 & 0.39 & \change{1.27} & 3.08      \\
						& \textbf{20}       & \textbf{2.09\change{$\pm$0.04}} & \textbf{1.75} & \textbf{0.34} & \textbf{\change{1.08}} & \textbf{3.16}      \\
            &          &      &      &      &      &  \\  

	2xCPU      & --       & 3.70\change{$\pm$0.03} & 1.00 & --   & 2.70 & --        \\
	Phi+2xCPU  & 1        & 2.94\change{$\pm$0.03} & 0.98 & 1.96 & \change{0.88} & 1.26      \\
						& 5        & 1.61\change{$\pm$0.04} & 1.01 & 0.5 & \change{0.99} & 2.30      \\
						& \textbf{10}       & \textbf{1.47\change{$\pm$0.05}} & \textbf{1.05} & \textbf{0.42} & \textbf{\change{1.03}} & \textbf{2.52}      \\
						& 15       & 1.59\change{$\pm$0.05} & 1.09 & 0.49 &\change{1.27}  & 2.33      \\
						& 20 & 1.52\change{$\pm$0.05} & 1.12 & 0.40  & \change{1.08} & 2.43      \\
&          &      &      &      &      &  \\

GPU+CPU    & 20       & 1.84 & 1.73 & 0.10 & 0.47 & 3.60      \\ 
GPU+2xCPU  & 20       & 1.22 & 1.12 & 0.10 & 0.47 & 3.03      \\ 
\hline \end{tabular}
\par\end{centering}
\begin{centering}
\bigskip{}
\par\end{centering}
\begin{centering}
\textbf{double precision}
\par\end{centering}
\centering{}\begin{tabular}{lllllll} \hline  Hardware   & slices   & W     & L    & O   & A   & speedup   \\
\hline
	CPU        & --     & 12.05\change{$\pm$0.01}  & 2.80 & --    & 9.26 & --  \\
	Phi+CPU    & 1        & 6.76\change{$\pm$0.03} & 2.84 & 3.92  & \change{1.80}  & 1.78   \\
						& 5        & 3.97\change{$\pm$0.06} & 2.79 & 1.18  & \change{2.04}  & 3.04   \\
						& 10       & 3.60\change{$\pm$0.07} & 2.78 & 0.82  &  \change{2.15} & 3.35   \\
						& 15       & 3.63\change{$\pm$0.10} & 2.86 & 0.78  & \change{2.73}  & 3.32    \\
			 & \textbf{20}       & \textbf{3.48\change{$\pm$0.09}}  & \textbf{2.86} & \textbf{0.62} & \textbf{\change{2.20}} & \textbf{3.46}  \\
            &          &       &      &      &      & \\
 
	2xCPU      & --       & 7.01\change{$\pm$0.01} & 1.91 & --    & 5.11 & --  \\
	Phi+2xCPU  & 1        & 5.87\change{$\pm$0.02} & 1.92 & 3.95  & \change{1.80}  & 1.20  \\
						& 5        & 3.13\change{$\pm$0.06} & 1.97 & 1.16  & \change{2.03}  & 2.24  \\
						& \textbf{10}       & \textbf{2.84\change{$\pm$0.08}} & \textbf{2.04} & \textbf{0.80} & \textbf{\change{2.15}}  & \textbf{2.47}  \\
						& 15       & 3.17\change{$\pm$0.07} & 2.07 & 1.10  & \change{2.77}  & 2.22  \\
						& 20       & 2.91\change{$\pm$0.09} & 2.12 & 0.79  & \change{2.17}  & 2.41  \\
            &          &       &      &      &      & \\

   GPU+CPU  & 15       & 2.99 & 2.77 & 0.22  & 0.78 & 4.03  \\
  GPU+2xCPU & 15       & 2.44 & 2.20 & 0.24  & 0.78 & 2.88  \\ 
\hline \end{tabular}
\end{table}
Note that the performance of the GPU implementation \change{on one-half of the NVIDIA K80} (considered in
section \ref{sec:single-gpu}) is superior by approximately 20\% (for
the dual socket case) and approximately 15\% (for the single socket
case) compared to the Xeon Phi \change{7120} implementation. We should also note
that, as discussed before, the interleave scheme is out of necessity
somewhat more complicated than the interleave scheme that is used
for the GPU code (see Figure \ref{fig:GPU-interleave1}).

The performance difference between the Intel Xeon Phi \change{7120} and the \change{NVIDIA K80}
are mainly explained by the fact that the assembly step is more costly
on the Xeon Phi \change{7120}. Therefore, it is not possible to hide the data transfer
as well as on the \change{K80} which negatively impacts the performance of
the implementation.

\section{Multiple GPU implementation\label{sec:multiple-gpu}}

The GPU implementation in section \ref{sec:single-gpu} uses a single
GPU to perform the assembly step of the optimization algorithm. However,
as has been pointed out in the introduction, the NVIDIA K80 includes
two identical GPUs within the same expansion card. Thus, so far we
have only used one half of the computational potential within that
package. Certainly, we can not expect a factor of two improvement
when using this additional GPU as in the present implementation performance
is mainly limited by the linear solve conducted on the CPU. However,
the timing results given in Table \ref{tab:cpu-vs-all} suggest that
we could solve part of the problem (both assembly and linear solve)
on the second GPU. In this situation, optimal load balancing dictates
the amount of work that is parceled out to the second GPU. Based on
Table \ref{tab:cpu-vs-all} we would expect that we achieve optimal
performance by assigning 35\% (double precision, single socket), 30\%
(double precision, dual socket and single precision, single socket),
and 20\% (single precision, dual socket) of the work set to the second
GPU. Thus, in most situations we would expect a maximal speedup of
about 40-50\% (compared to the single GPU implementation). The exception
being the single precision dual socket configuration in which a maximal
speedup of only 25\% is possible.

\chgtwo{Since both the assembly step and
the linear solve are computed on the second GPU, we first completely assemble the systems (using a single CUDA kernel call)
and then perform the linear solves (using a single MAGMA call). In this process no data needs to be
transferred to or from the GPU and we do not divide our parcel of the workload into slices. In fact, doing the
latter incurs a small but significant performance penalty.}

There is one additional issue that deserves our attention. While the
MAGMA linear algebra library includes routines that use the GPU memory
as input and output, it is primarily designed to operate in an environment
that includes CPUs as well as GPUs. Consequently, there is no way
to execute a MAGMA routine without CPU support and in an asynchronous
fashion. To avoid oversubscription (which measurements show has a
negative impact on performance) we use only 15 OpenMP threads for
the linear solve and execute the MAGMA call in a separate pthread.
However, it is clear that this reduces the maximal achievable improvement
in performance to a certain degree.

The timing results for this implementation are shown in Table \ref{tab:timing-2xgpu}.
We observe a speedup of $3.4$ (single precision) and $3.8$ (double
precision) for adding a K80 to the dual socket workstation. In the
case of a single socket workstation the observed speedup is approximately
$4.7$ (single precision) and $5.6$ (double precision). We remark
that the speedup compared to the single GPU implementation is in all
cases within 5\% of the maximal achievable speedup (based on the design
decisions outlined in this section).

\begin{table}[h]
\caption{Timing results for the hybrid algorithm (\change{full NVIDIA K80}+CPU) that uses both
GPUs of the K80 hardware. The wall time (W), the time required to
assemble the system (A), the time required for the linear solves (L)
and the overhead due to offloading to the GPUs (O) is shown.
\change{Note that for the GPU implementation the time required by the linear solve (which is done on the CPU) always dominates the total runtime. Thus, we have $\text{W} = \text{L}+\text{O}$.}
The number of slices and work distribution that yields the optimal run time are
highlighted in bold in the table. All measured times are in units
of seconds. \change{In addition, the standard deviation determined from $20$ repetitions of the simulation is shown next to the wall time.} \label{tab:timing-2xgpu}}

\begin{centering}
\textbf{single precision}
\par\end{centering}
\begin{centering}
\begin{tabular}{lllllll} \hline  Hardware   & slices,distr   & W    & L    & O   & A     & speedup   \\
\hline
	CPU         & --      & 6.61\change{$\pm$0.01} & 1.68 & --   & 4.93  & --        \\
	2xGPU+CPU   & 15,0.35 & 1.57\change{$\pm$0.01} & 1.26 & 0.31 & 0.45  & 4.21  \\ 
						 & \textbf{15,0.30}  & \textbf{1.41\change{$\pm$0.02} } & \textbf{1.28} & \textbf{0.13} & \textbf{0.42} & \textbf{4.68}  \\
						 & 20,0.30 & 1.42\change{$\pm$0.02} & 1.29 & 0.13  & 0.42 & 4.65   \\
						 & 20,0.25 & 1.48\change{$\pm$0.02} & 1.36 & 0.12  & 0.35 & 4.47  \\ 
             &          &      &      &      &      &         \\
 
	2xCPU       & --       & 3.70\change{$\pm$0.03} & 1.00 & --   & 2.70 & --  \\ 
	2xGPU+2xCPU & 15,0.30  & 1.38\change{$\pm$0.01} & 0.92 & 0.46 & 0.33 & 2.68 \\ 
						 & 15,0.25  & 1.18\change{$\pm$0.01} & 0.95 & 0.23 & 0.35 & 3.14 \\ 
						 & \textbf{15,0.20} & \textbf{1.08\change{$\pm$0.02} }& \textbf{0.98} & \textbf{0.11} & \textbf{0.37}  & \textbf{3.41} \\
						 & 20,0.20  & 1.10\change{$\pm$0.02} & 0.99 & 0.11& 0.37  & 3.37 \\ 
			 &          &      &      &      &      &         \\

GPU+CPU     & 20       & 1.84& 1.73 & 0.10 & 0.47  & 3.60   \\ 
GPU+2xCPU   & 20       & 1.22& 1.12 & 0.10 & 0.47  & 3.03    \\ 
\hline \end{tabular}
\par\end{centering}
\begin{centering}
\bigskip{}
\par\end{centering}
\begin{centering}
\textbf{double precision}
\par\end{centering}
\centering{}\begin{tabular}{lllllll} \hline  Hardware   & slices,distr   & W    & L    & O    & A     & speedup   \\
\hline
	CPU        & --      & 12.05\change{$\pm$0.01} & 2.80 & --   & 9.26  & --   \\
	2xGPU+CPU  & 10,0.40  & 2.31\change{$\pm$0.01} & 1.79 & 0.52 & 0.47  & 5.23  \\
						& \textbf{10,0.35} & \textbf{2.14\change{$\pm$0.04} } & \textbf{1.91} & \textbf{0.23} & \textbf{0.51}  & \textbf{5.63}   \\ 
						& 15,0.35 & 2.15\change{$\pm$0.03 } & 1.94 & 0.22 & 0.51 & 5.60  \\
						& 15,0.30  & 2.30\change{$\pm$0.03} & 2.09 & 0.21 & 0.55 & 5.23 \\ 
            &          &       &      &      &      &   \\
     
	2xCPU       & --      & 7.01\change{$\pm$0.01} & 1.91 & --   & 5.11  & --  \\
	2xGPU+2xCPU & 10,0.35 & 2.08\change{$\pm$0.01} & 1.50 & 0.58 & 0.51  & 3.38 \\ 
						 & 10,0.30  & 1.88\change{$\pm$0.01} & 1.64 & 0.23 & 0.55  & 3.73 \\ 
						 & \textbf{15,0.30}  & \textbf{1.86\change{$\pm$0.02} }  & \textbf{1.63} & \textbf{0.23} & \textbf{0.55} & \textbf{3.78} \\
						 & 15,0.25 & 1.90\change{$\pm$0.02} & 1.67 & 0.23 & 0.59  & 3.70 \\ 
             &          &       &      &      &      &   \\

GPU+CPU   & 20       & 3.08 & 2.88 & 0.20 & 0.78  & 3.91 \\
GPU+2xCPU & 15       & 2.44 & 2.20 & 0.24  & 0.78 & 2.88 \\ 
\hline \end{tabular}

\end{table}

\section{Conclusion\label{sec:conclusion}}

We have compared the speedup that can be achieved for a genetic optimization
algorithm that uses a panel method as the inner solver when an Intel
Xeon Phi 7120 or a NVIDIA K80 is added to a workstation with one or
two Intel Xeon E5-2630 v3 CPUs. Optimization and parallelization for
the CPU and Intel Xeon Phi code is done using the Intel C compiler
(vectorization) and OpenMP. For the GPU we use an implementation that
is based on CUDA. Since the linear solver is faster on the CPU and
the assembly is faster on the \change{Xeon Phi 7120/NVIDIA K80}, the present algorithms
profits from a hybrid implementation that uses both traditional CPUs
as well as accelerators. The obtained results can be summarized as
follows:
\begin{itemize}
\item Adding a K80 to the dual socket workstation results in a speedup of
approximately $3.4$ (single precision) and $3.8$ (double precision).
\item Adding a Xeon Phi 7120 to the dual socket workstation results in a
speedup of approximately $2.4$ (single precision) and $2.5$ (double
precision).
\item Since the performance of the CPU only implementation is mostly dominated
by the assembly step, the speedups for a single CPU are significantly
larger. In this configuration we observe speedups of up to $5.6$
        on the \change{NVIDIA K80} and up to $3.5$ for the Xeon Phi \change{7120} implementation.
\end{itemize}
These speedups are clearly of practical interest. This is true both
for the \change{NVIDIA K80} as well as for the Xeon Phi \change{7120}. For the problem
under consideration the \change{NVIDIA K80} yields better performance
compared to the Xeon Phi \change{7120}. What is not so clear cut is the development
effort that is required for each platform. One advantage of the Xeon
Phi is that once we had an optimized code for the assembly step on
the CPU (using vectorization and OpenMP) we almost immediately obtained
good performance on the Xeon Phi. On the other hand, the CUDA implementation
of the assembly step is straightforward and due to the computational
advantage of the GPU a less complicated communication hiding scheme
proves sufficient. Thus, with respect to the development effort involved
there is no clear winner.

\bibliography{panelmethods}

\begin{thebibliography}{10}

\bibitem{casas2014}
Cas{\'a}s VD, Duro RJ, Lopez-Pena F.
\newblock {Evolutionary design of wind turbine blades}.
\newblock International Journal of Computing. 2014;4(3):49--55.

\bibitem{jones2000}
Jones BR, Crossley WA, Lyrintzis AS.
\newblock {Aerodynamic and aeroacoustic optimization of rotorcraft airfoils via
  a parallel genetic algorithm}.
\newblock Journal of Aircraft. 2000;37(6):1088--1096.

\bibitem{gardner2003}
Gardner BA, Selig MS.
\newblock {Airfoil design using a genetic algorithm and an inverse method}.
\newblock In: 41st Aerospace Sciences Meeting and Exhibit; 2003. p. 1--12.

\bibitem{drela1989}
Drela M.
\newblock {XFOIL: An Analysis and Design System for Low Reynolds Number
  Airfoils}.
\newblock In: Mueller TJ, editor. {Low Reynolds Number Aerodynamics}. vol.~54
  of Lecture Notes in Engineering. Springer Berlin Heidelberg; 1989. p. 1--12.

\bibitem{kipouros2012}
Kipouros T, Peachey T, Abramson D, Savill AM.
\newblock {Enhancing and developing the practical optimisation capabilities and
  intelligence of automatic design software}.
\newblock In: 8th AIAA Multi-Disciplinary Design Optimization Specialist
  Conference, Honolulu, Hawaii; 2012. p. 1--7.

\bibitem{gabor2012l}
Gabor OS, Koreanschi A, Botez RM.
\newblock {Low-speed aerodynamic characteristics improvement of ATR 42 airfoil
  using a morphing wing approach}.
\newblock In: IECON 2012--38th Annual Conference on IEEE Industrial Electronics
  Society. IEEE; 2012. p. 5451--5456.

\bibitem{fincham2015}
Fincham JHS, Friswell MI.
\newblock {Aerodynamic optimisation of a camber morphing aerofoil}.
\newblock Aerospace Science and Technology. 2015;43:245--255.

\bibitem{morgado2016}
Morgado J, Vizinho R, Silvestre MAR, P{\'a}scoa JC.
\newblock {XFOIL vs CFD performance predictions for high lift low Reynolds
  number airfoils}.
\newblock Aerospace Science and Technology. 2016;52:207--214.

\bibitem{novikov2014}
T\"urkal M, Novikov Y, \"Usenmez S, Sezer-Uzol N, Uzol O.
\newblock {GPU based fast free-wake calculations for multiple horizontal axis
  wind turbine rotors}.
\newblock In: Journal of Physics: Conference Series. vol. 524. IOP Publishing;
  2014. p. 012100.

\bibitem{chabalko2014}
Chabalko CC, Balachandran B.
\newblock {Implementation and benchmarking of two-dimensional vortex
  interactions on a graphics processing unit}.
\newblock Journal of Aerospace Information Systems. 2014;11(6):372--385.

\bibitem{morgenthal2014}
Morgenthal G, Corriols AS, Bendig B.
\newblock {A GPU-accelerated pseudo-3D vortex method for aerodynamic analysis}.
\newblock Journal of Wind Engineering and Industrial Aerodynamics.
  2014;125:69--80.

\bibitem{develder2014}
deVelder NB.
\newblock {Free Wake Potential Flow Vortex Wind Turbine Modeling: Advances in
  Parallel Processing and Integration of Ground Effects}.
\newblock University of Massachusetts-Amherst; 2014.

\bibitem{chabalko2013}
Chabalko C, Fitzgerald T, Balachandran B.
\newblock {GPGPU implementation and benchmarking of the unsteady vortex lattice
  method}.
\newblock In: 51st AIAA aerospace sciences meeting.; 2013. .

\bibitem{tavoularis}
{Tavoularis, S}. {Thwaites{'} Method for Laminar Boundary Layers with Pressure
  Gradient (Lecture notes)};.
\newblock
  \url{http://by.genie.uottawa.ca/~mcg4345/AdditionalNotes/45_Thwaites.pdf}.

\bibitem{moran1984}
Moran J.
\newblock {An introduction to theoretical and computational aerodynamics}.
\newblock Dover; 2003.

\bibitem{nordin1989}
Nordin A, Wah LC.
\newblock {Thwaites' Method In Laminar Boundary Layer}.
\newblock Jurnal Teknologi. 1989;14(1):5--13.

\bibitem{rogalsky2000}
Rogalsky T, Kocabiyik S, Derksen RW.
\newblock {Differential evolution in aerodynamic optimization}.
\newblock Canadian Aeronautics and Space Journal. 2000;46(4):183--190.

\bibitem{poli2008}
Poli R, Langdon WB, McPhee NF, Koza JR.
\newblock {A field guide to genetic programming}.
\newblock Lulu Enterprises; 2008.

\bibitem{dong2014}
Dong T, Haidar A, Luszczek P, Harris JA, Tomov S, Dongarra J.
\newblock {LU factorization of small matrices: accelerating batched DGETRF on
  the GPU}.
\newblock In: High Performance Computing and Communications. IEEE; 2014. p.
  157--160.

\bibitem{vladimirov2015}
Vladimirov A.
\newblock {Fine-Tuning Vectorization and Memory Traffic on Intel Xeon Phi
  Coprocessors: LU Decomposition of Small Matrices}.
\newblock Colfax Research. 2015;p. 1--10.

\end{thebibliography}

\end{document}